\documentclass[a4paper,12pt]{scrartcl}

\usepackage{cite}
\usepackage{color}
\usepackage{xspace}
\usepackage{graphicx}
\usepackage{pslatex}
\usepackage{amsmath}
\usepackage{txfonts}
\usepackage[pdfpagelabels]{hyperref}
\usepackage[utf8]{inputenc}

\usepackage{listings}

\graphicspath{{figs/}}

\newcommand*{\mr}{\mathrm}
\newcommand*{\bm}{\boldsymbol}
\newcommand{\del}{\partial}                     
\newcommand*{\kira}{\texttt{Kira}}
\newcommand*{\pyred}{\texttt{pyRed}}
\newcommand*{\ice}{\texttt{ICE}}
\newcommand*{\cplusplus}{\texttt{C++}}
\newcommand*{\ginac}{\texttt{GiNaC}}
\newcommand*{\cln}{\texttt{CLN}}
\newcommand*{\zlib}{\texttt{zlib}}
\newcommand*{\yaml}{\texttt{yaml}}
\newcommand*{\yamlcpp}{\texttt{yaml-cpp}}
\newcommand*{\air}{\texttt{AIR}}
\newcommand*{\fire}{\texttt{FIRE}}
\newcommand*{\firefive}{\texttt{FIRE\,5}}
\newcommand*{\reduze}{\texttt{Reduze}}
\newcommand*{\reduzetwo}{\texttt{Reduze\,2}}
\newcommand*{\gatetofermat}{\texttt{gateToFermat}}

\newcommand*{\fermat}{\texttt{Fermat}}
\newcommand*{\sqlite}{\texttt{SQLite3}}
\newcommand*{\python}{\texttt{Python}}
\newcommand*{\form}{\texttt{FORM}}
\newcommand*{\mathematica}{\texttt{Mathematica}}

\newcommand*{\CE}{\mathcal{E}}

\newcommand*{\CG}{\mathcal{G}}
\newcommand*{\CI}{\mathcal{I}}

\def\Eq#1{{Eq.~(\ref{#1})}}
\def\Fig#1{{Fig.~\ref{#1}}}
\def\Tab#1{{Tab.~\ref{#1}}}
\titlehead{
{\flushright{
\small HU-EP-17/08
\\
\small FR-PHENO-2017-14
\\
\small TCDMATH 18--02
\\[8mm]
}}
}

\title{\Large Kira -- A Feynman Integral Reduction Program}
\author{\large  P. Maierh\"ofer$\,^a$, J. Usovitsch${^b}{^c}$  and P.~Uwer$\,^b$ \\
  \large $^a$ \textit{Physikalisches Institut, Albert-Ludwigs-Universit\"at
    Freiburg},\\
  \large\textit{79104~Freiburg, Germany}\\
  \large $^b$\textit{Humboldt-Universit\"at zu Berlin, Institut f\"ur
    Physik},\\
  \large \textit{Newtonstra{\ss}e~1:5, 12489~Berlin, Germany}\\  
  \large $^c$\textit{Hamilton Mathematics Institute, Trinity College Dublin},\\
  \large \textit{College~Green, Dublin~2, Ireland}  
  }
\date{}

\begin{document}
\maketitle
\begin{abstract}
  In this article, we present a new implementation of the Laporta
  algorithm to reduce scalar multi-loop integrals---appearing in
  quantum field theoretic calcula\-tions---to a set of master
  integrals. We extend existing approaches by using an additional
  algorithm based on modular arithmetic to remove linearly dependent
  equations from the system of equations arising from
  integration-by-parts and Lorentz identities. Furthermore, the
  algebraic manipulations required in the back substitution are
  optimized. We describe in detail the implementation as well as the
  usage of the program. In addition, we show benchmarks for concrete
  examples and compare the performance to \reduzetwo{} and \firefive.
  In our benchmarks we find that \kira{} is highly competitive with these
  existing tools.
\end{abstract}

\newpage

\textbf{PROGRAM SUMMARY}

\vspace{1cm}

\begin{small}
\noindent
{\em Manuscript Title:} Kira -- A Feynman Integral Reduction Program\\
{\em Authors:} P. Maierh\"ofer, J. Usovitsch  and P. Uwer\\
{\em Program title:} \kira\\
{\em Licensing provisions:} GPLv3 or later\\
{\em Programming language:} \texttt{C++}\\
{\em Computer(s) for which the program has been designed:} desktop PC, compute nodes/workstations\\
{\em Operating system(s) for which the program has been designed:} Linux 64bit\\
{\em RAM required to execute with typical data:} depends on the complexity of the problem, from few MB up to a few hundred GB, or even
    more in complicated cases.\\
{\em Has the code been vectorized or parallelized?:} yes\\
{\em Number of processors used: } any number of cores\\
{\em Supplementary material:} this article, examples
{\em CPC Library Classification:} 4.4 Feynman diagrams, 4.8 Linear Equations and Matrices, 5 Computer Algebra\\
{\em External routines/libraries used:} \fermat{} [1], \gatetofermat{} [2], \ginac{} [3,4], \yamlcpp{} [5], \zlib{} [6] and \sqlite{} [7]\\
{\em Nature of problem:}
The reduction of Feynman integrals to master integrals leads in
general to a system of equations which contains redundant, i.e.\ linearly
dependent, equations.  In particular, for multi-scale problems, the algebraic
manipulation of these redundant equations can lead to a substantial increase in
runtime and memory consumption without affecting the results. \\
{\em Solution method:}
The program identifies linearly dependent relations based on modular
arithmetic with the help of an algorithm presented in Ref.~[8]. Afterwards the program brings a linearly independent system of equations in a triangular form. Furthermore, the algebraic  manipulations required in the back substitution are optimized.\\
{\em Restrictions:} the CPU time and the available RAM\\
{\em Running time:} minutes to weeks, depending on the complexity of the problem\\
{\em References:} 
{\\} [1] R. H. Lewis, \textit{Computer Algebra System Fermat}, \url{https://home.bway.net/lewis/}.
{\\} [2] M. Tentioukov, \textit{gateToFermat}, \url{http://science.sander.su/FLink.htm}.
{\\} [3] C. W. Bauer, A. Frink, and R. Kreckel, \textit{Introduction to the GiNaC framework for symbolic computation within the C++ programming language}, J. Symb. Comput. \textbf{33} (2000) 1, \href{http://arxiv.org/abs/cs/0004015}{{\ttfamily arXiv:cs/0004015 [cs-sc]}}.
{\\} [4] J. Vollinga, \textit{GiNaC: Symbolic computation with C++}, Nucl. Instrum. Meth. \textbf{A559} (2006) 282-284, \href{http://arxiv.org/abs/hep-ph/0510057}{{\ttfamily arXiv:hep-ph/0510057 [hep-ph]}}.
{\\} [5] YAML, \textit{YAML Ain't Markup Language}, \url{http://yaml.org}.
{\\} [6] J.-L. Gailly and M. Adler, \textit{ZLIB}, \url{http://zlib.net}.
{\\} [7] SQLite, \textit{SQLite3, version: 3.14.2}, \url{https://www.sqlite.org}.
{\\} [8] P. Kant, \textit{Finding Linear Dependencies in Integration-By-Parts Equations: A Monte Carlo Approach}, Comput. Phys. Commun. \textbf{185} (2014) 1473-1476, \href{http://arxiv.org/abs/1309.7287}{{\ttfamily arXiv:1309.7287 [hep-ph]}}.
\end{small}

\newpage

\section{Introduction}

The steadily increasing experimental precision reached in current collider
experiments like ATLAS and CMS requires on the theory side the evaluation of
higher order corrections in the perturbative expansion. While the computation of
next-to-leading order (NLO) corrections is well established today, the same
level of maturity has not yet been reached for next-to-next-to-leading order
(NNLO) calculations, although tremendous progress has been made in the last few
years, see for example
\cite{Czakon:2013goa,Boughezal:2015dra,Chen:2014gva,Lindert:2017pky,%
Borowka:2016ehy,Ridder:2015dxa,Boughezal:2015ded,Boughezal:2015dva,%
Grazzini:2016swo,Grazzini:2015nwa,Gehrmann:2014fva,Caola:2015rqy,%
Caola:2015ila,Cascioli:2014yka,Caola:2015psa,Currie:2017tpe,Anastasiou:2016cez}
for an incomplete list of recent calculations.

One major bottleneck in the evaluation of multi-loop amplitudes is the
computation of the occurring Feynman integrals. The application of
Feynman rules leads in general to tensor integrals. All Feynman
integrals may be calculated directly. This approach is worked out in
\cite{Dubovyk:2016aqv,Dubovyk:2016ocz,Dubovyk:2016zok}. However, in
most applications it is advantageous to reduce the tensor integrals to
scalar integrals.  Because of integration-by-parts
\cite{Tkachov:1981wb,Chetyrkin:1981qh} and Lorentz
\cite{Gehrmann:1999as} identities these integrals are not independent
and can be expressed in terms of a small set of so-called master
integrals.  The integration-by-parts and Lorentz identities relate
integrals with different powers of the propagators. Combining these
relations algebraically, `ladder-operators' to reduce the powers of
the propagators in form of a recursion can be constructed.  In
practice, this procedure is however highly non-trivial and cumbersome.

Alternatively, the integration-by-parts and Lorentz relations can be evaluated
for integer (instead of algebraic) powers of the propagators. Using different
integer values for the different powers as seeds, a system of equations can be
set up. Solving this system leads to a reduction to master
integrals. This is the essence of the Laporta algorithm \cite{Laporta:2001dd}.
Since the integral reduction is a crucial step in the analytic evaluation of
multi-loop amplitudes, various publicly available implementations of the Laporta
algorithm exist: \air{} \cite{Anastasiou:2004vj}, \fire{} \cite{Smirnov:2008iw,Smirnov:2013dia,Smirnov:2014hma}
and \reduze{}~\cite{Studerus:2009ye,vonManteuffel:2012np}. Applying these
programs to state-of-the-art calculations depending on several mass scales
(internal/external particle masses and scalar products of external momenta) the
required runtime and the memory consumption may put severe limits in practical
applications.

In this article we present an optimized implementation of the Laporta algorithm
with the aim to extend the frontier of achievable reductions to more mass
scales. Generating the system of equations using different input seeds leads in
general to a system of equations which contains redundant, i.e.\ linearly
dependent, equations.  In particular, for multi-scale problems, the algebraic
manipulation of these redundant equations can lead to a substantial increase in
runtime and memory consumption without affecting the results. In
Ref.~\cite{Kant:2013vta} a method has been presented to eliminate the linearly
dependent equations using only fixed-size integer arithmetic instead of
computationally intense algebraic manipulations. The main idea is to replace the
different mass scales occurring in the problem by integer numbers over a finite
field and perform a Gauss type elimination afterwards to identify dependent
equations. Besides the elimination of redundant equations, this procedure allows
us to identify the master integrals before or even without performing the actual
reduction, a task for which otherwise dedicated algorithms or computer
programs are required, e.g.~\cite{Lee:2013hzt,Georgoudis:2016wff}. Furthermore, the
handling of the algebraic integral coefficients occurring in the reduction
tables is improved. We find that these modifications lead to a substantial
improvement in performance, in particular, when multi-scale problems are
studied. In addition, since \kira{} uses input very similar to the one required
by \reduzetwo{}, our implementation can also be used to perform independent cross
checks of results generated with \reduzetwo{}.

The outline of this article is as follows. To introduce the notation
we briefly review in section 2 some basic aspects of multi-loop
Feynman integrals. In section 3 we describe our implementation of the
Laporta algorithm. Section 4 gives detailed information on the
required prerequisites and how to install \kira{}. In section 5 we
illustrate the usage with a simple example. In addition, information
on various options to tune the reduction is provided.  Section 6
presents some benchmarks. More precisely, three double box topologies
with non-vanishing internal and external masses are reduced and the
required runtime is reported. As reference we also present the runtime
required using the program \reduzetwo{}
\cite{Studerus:2009ye,vonManteuffel:2012np}. We finally close with a
conclusion in section 7.

\section{Preliminaries }
\label{feynman integral}

To fix the notation used in this work we start with a brief review of
multi-loop integrals as encountered in perturbative calculations
in quantum field theory.  Applying within a concrete model the
Feynman rules to calculate scattering matrix elements
leads to multi-loop tensor integrals of the form
\begin{equation}
  \int \prod_{i=1}^L d^d\ell_i\;
  \frac{\ell_1^{\mu_1}\cdots \ell_1^{\mu_j}\ldots
        \ell_L^{\nu_1}\cdots \ell_L^{\nu_m}}
       {P_1(\ell_1,\ldots,\ell_L,p_1,\ldots p_E) \cdots
        P_{t'}(\ell_1,\ldots,\ell_L,p_1,\ldots p_E)}.
\end{equation}
The $E$ momenta $p_i$ denote the linearly independent external
momenta. (We consider a scattering amplitude with $E+1$ external
momenta/legs, however, because of momentum conservation only $E$
momenta are independent.)  The $L$ momenta $\ell_i$ are the loop
momenta which are not fixed through momentum conservation at each
vertex. With $t'$ we denote the number of propagators of which $t$ are
independent.  The inverse propagators $P_i$ are of the form
\begin{equation}
  P_i = k_i^2 - m_i^2 + i\varepsilon,
\end{equation}
with $k_i$ being a linear combination of the momenta
$\ell_1,\ldots,\ell_L$ and $p_1,\ldots, p_E$ and $m_i$ denoting the
masses of the corresponding virtual particles. Within dimensional
regularization $d=4-2\epsilon$ denotes the dimension of space-time. As
usual $d\not=4$ is used to regularize infra-red and ultraviolet
divergences.  Using projectors or a Feynman/Schwinger type
parametrization the multi-loop tensor integrals can be reduced to
scalar multi-loop integrals. The Feynman/Schwinger type
parametrization will introduce scalar integrals with shifted
dimensions and indices. The projectors will generate scalar integrals
with auxiliary propagators which represent irreducible scalar
propagators.
The required number of auxiliary propagators is
easily calculated. The number of scalar products involving the
loop-momenta is given by
\begin{equation}
  N = EL + \frac{L(L+1)}{2}.
\end{equation}
However, $t$ scalar products can be expressed in terms of linear combinations
of the propagators. The number of auxiliary propagators is thus given
by $(N-t)$. The occurring integrals can thus be cast in the form
\begin{equation}
  \label{eq:NotationIntegrals}
  T(d,a_1,\ldots,a_t,a_{t+1},\ldots, a_{N},\{p_j\}) =
  \int\prod_{i=1}^L d^d\ell_i\;
  \frac{1}{P_1^{a_1} \ldots P_t^{a_t} P_{t+1}^{a_{t+1}}\ldots P_N^{a_{N}}},
\end{equation}
where the powers $a_i$ of the auxiliary propagators (i.e.\ $i=t+1\ldots N$)
may only take non-positive values. Note that the
auxiliary propagators $P_{t+1}\ldots P_N$ are not uniquely fixed. They
are constrained only by the requirement that together with the first $t$
propagators all scalar products involving the loop momenta are
expressible as linear combinations of the $N$ propagators. As a short 
hand notation we collect the indices $a_i$ into an $N$
dimensional vector $\bm{a}=(a_1,\ldots,a_N)$.

\paragraph{Integration-by-parts and Lorentz-invariance identities:}
Performing the reduction of the tensor integrals to scalar integrals
outlined above leads in general to a large number of scalar integrals.
However, these integrals are not independent. So called \textit{
  Integration-By-Parts} (IBP) identities
\cite{Tkachov:1981wb,Chetyrkin:1981qh} and \textit{
  Lorentz-Invariance} (LI) \cite{Gehrmann:1999as} lead to linear
relations between them. As a consequence the large number of scalar
integrals can be reduced to a smaller set of master integrals, which
serve as a basis to express all other scalar integrals. To be more
specific the IBP equations follow from
\begin{equation}
  \int\prod\limits^{L}_{j=1}d^{d}\ell_{j}
  \frac{\partial}{\partial\ell_f^\mu}
  \left(\frac{q_l^\mu}{P_1^{a_1} \ldots P_N^{a_{N}}}\right) = 0,
  \quad f=1,\dots,L, \quad l=1,\dots,L+E,
  \label{eq:Definition IBP2}
\end{equation}
with $q_l = \ell_l$ for $l = 1\ldots L$ and $q_l = p_{l-L}$ for
$l=L+1\ldots L+E$. For a fixed vector $\bm{a}$ the possible choices
for $f$ and $l$ lead to $L(E+L)$ IBP equations relating integrals with
indices shifted by one unit to each other.

The LI equations follow from
\begin{equation}
  \sum\limits_{i=1}^{E}\left(p^{\nu}_{i}\frac{\del}{\del
      p_{i\mu}}-p_{i}^{\mu}\frac{\del}{\del
      p_{i\nu}}\right)T(\bm{a},\{p_{i}\})=0.
\label{eq:lorentz}
\end{equation}
Contracting this equation with all possible antisymmetric combinations
of the form
\begin{equation}
  p_{r\mu}p_{s\nu}-p_{s\mu}p_{r\nu},
\end{equation}
leads to $E(E-1)/2$ equations between integrals with shifted indices.
To reduce the large number of scalar integrals to the master integrals
there are essentially two different strategies. One method is to
combine the LI and IBP relations to construct `ladder-operators' for
the individual propagators. A recursive application of these 
ladder-operators can then be used to reduce all integrals to the master
integrals. However, in practice this approach suffers from the fact
that the construction of the ladder-operators is non-trivial and
often involves some handwork. For recent progress in this direction we
refer to Refs.~\cite{Smirnov:2013dia,Lee:2013mka,Ruijl:2017cxj}. In the second
approach the IBP- and LI-equations are applied to integer $\bm{a}\in
\mathbb{Z}^N$ instead of algebraic $\bm{a}$. Making different choices
for $\bm{a}$ which are often called \textit{seeds} a huge system of
equations can be built up.  Using different seeds leads in general to
relations between different (unknown) scalar integrals. However, it
turns out that the number of relations grows faster than the number
of unknown integrals. Making the system big
enough all required scalar integrals can be reduced to the master
integrals by applying a Gauss type elimination algorithm. This is the
essence of the Laporta approach first described in Ref.~\cite{Laporta:2001dd}.
Alternative ways of using IBP identities for integral reduction have been 
explored e.g. in \cite{Ita:2015tya,Larsen:2015ped}

\paragraph{Sectors and sub-sectors:} In practical applications it turns
out that the system of equations typically shows some block structure.
Since respecting this structure in the reduction may be beneficial it
is useful to introduce the notion of sectors and sub-sectors.  For a
given scalar integral the related sector is defined as the set of
integrals for which the subset of propagators occurring with positive
powers is the same. For each $\bm{a}$ we define a vector $\bm{\theta}
= (\theta_1,\ldots,\theta_N)$ where the $\theta_i$ are set to one if
$a_i>0$ and zero otherwise,
\begin{equation}
  \theta_i = \Theta\big(a_i-\tfrac{1}{2}\big),
\end{equation}
where $\Theta(x)$ is the Heaviside step function.
All scalar integrals within one sector lead to the same
$\bm{\theta}$. The scalar integral for which $\bm{a_C} =
\bm{\theta}(\bm{a_C})$ is called the corner integral of the sector.
To uniquely label a sector we may identify the $\theta_i$ as the
components of a binary representation of a sector id $S$,
\begin{equation}
  \label{eq:sectornumber}
  S = \sum_{j=1}^N  \theta_j \cdot 2^{j-1}.
\end{equation}
The total number of possible sectors is given by $2^N$. The number of different
propagators appearing in the denominator of any integral of a sector
is given by
\begin{equation}
  \label{eq:t-definition}
  t = \sum_{i=1}^N \theta_i.
\end{equation}
Furthermore, we define the vector of positive propagator powers,
$\bm{r}=(r_{1},\dots,r_{t})$, which is obtained from $\bm{a}$ by removing
all non-positive indices (preserving the order), and analogously the vector
of negative propagator powers, $\bm{s}=(s_{1},\dots,s_{N-t})$.
Within a sector the sum of all positive powers of the propagators and the
negative sum of all non-positive powers constitute a measure for the complexity
of an integral. It is thus convenient to define
\begin{equation}
  \label{eq:rs-value}
  r = \sum_{i=1}^{t} r_i\quad\text{and}\quad s = -\sum_{i=1}^{N-t} s_i.
\end{equation}

\paragraph{Identification of trivial sectors:}
Within dimensional regularization, scaleless integrals are consistently set to
zero. In Ref.~\cite{Lee:2012cn} it is shown that if the corner integral of a
given sector is zero, all other integrals in this sector are zero, too.
Accordingly, such a sector is called a \textit{trivial sector} or a \textit{zero
sector}. To identify zero sectors, we employ the algorithm presented in
Ref.~\cite{Lee:2013mka}. The algorithm is based on the Feynman parameter
representation of Feynman integrals,
\begin{equation}
  T(d,\bm{a}) =
  \frac{\Gamma(a - L\frac{d}{2})}{\prod_i\Gamma(a_i)}
  \int \prod_{j=1}^{N} dz_j z_j^{a_j-1}\delta(1-z)
  \frac{\mathcal{F}^{\frac{d}{2}  L-a}}{\mathcal{U}^{{\frac{d}{2}}(L+1) - a}},
\end{equation}
with $z=\sum_{j=1}^N z_j$, $a=\sum_{j=1}^N a_j$ and the Symanzik polynomials
$\mathcal{U}$ and $\mathcal{F}$, which are multivariate polynomials in the
$z_j$. To identify zero sectors, the function
\begin{equation}
  G(\bm{z}) = \mathcal{F}(\bm{z}) + \mathcal{U}(\bm{z})
  \label{eq:G-polynom}
\end{equation}
is considered. If the equation
\begin{equation}
  \sum_i k_i z_i \frac{\partial G(\bm{z})}{ \partial z_i} = G(\bm{z})
\end{equation}
has a $z$-independent solution for $k_i$ the corresponding sector is trivial.
Identifying zero sectors in an early stage of the reduction procedure can
greatly simplify the reduction.

\paragraph{Symmetry relations between integrals:}
Another class of relations between Feynman integrals which are usually not
covered by IBP and LI identities is given by symmetry relations. A simple
example which exhibits such a symmetry is the one-loop bubble integral
\begin{equation}
  T(d,a_1,a_2) = \int d^d\ell \frac{1}{(\ell^2-m^2+i\varepsilon)^{a_1}
    ((\ell+p)^2-m^2+i\varepsilon)^{a_2}},
\end{equation}
which obeys the symmetry relation
\begin{equation}
  T(d,a_1,a_2) = T(d,a_2,a_1),
\end{equation}
corresponding to the shift
\begin{equation}
  \ell\to -\ell-p
\end{equation}
of the loop momentum $\ell$. In general, symmetries can be derived from loop
momentum shifts
\begin{equation}
  \ell'_i = \sum_{j=1}^{L} M_{ij} \ell_j + \sum_{j=1}^E c^{(i)}_j p_j
  \quad (i=1\ldots L),\quad M_{ij},c^{(i)}_j\in\{-1,0,1\},
\end{equation}
and may also involve permutations of external momenta which leave the Mandelstam
variables invariant. Such a transformation conveys a symmetry if applying it to
an integral with only non-negative powers $T(d,\bm{r})$ results in an integral
$T'(d,\bm{r}')$ where $\bm{r}'$ is a permutation of $\bm{r}$. Applying it to an
integral which contains negative powers results in a linear combination of
integrals. By employing such symmetry relations the number of independent
integrals can be reduced, resulting in a smaller set of master integrals.
Symmetries which relate integrals within the same sector to each other are
commonly referred to as sector symmetries. Those which relate different sectors of
the same or of different topologies to each other are referred to as sector
mappings. Furthermore, in certain cases symmetries exist which apply only to
integrals without negative propagator powers, because the relation can not be
represented in terms of loop momentum shifts and external momentum permutations.

The sector mappings and sector symmetries are identified by applying the 
equation~\eqref{eq:G-polynom} to each corner integral. If the function in 
equation~\eqref{eq:G-polynom} is equal for two different corner integrals after a 
permutation of Feynman parameters and kinematic invariants then 
the two considered corner integrals exhibit a symmetry relation described above.

\kira{} can handle several topologies in a single run. By exploiting mappings
between equivalent sectors of different topologies, a common set of linearly
independent master integrals for the entire set of topologies will be found.

\section{Laporta Algorithm -- Implementation}
\label{implementation}

In the Laporta algorithm the IBP, LI and symmetry relations are applied to a
chosen set of integrals $T(d,\bm{a})$ as defined in \Eq{eq:NotationIntegrals} to
generate a linear homogeneous system of equations $\CG$ with the integrals as
unknowns \cite{Laporta:2001dd}. In the implementation presented in this article
we use the \cplusplus{} library \ginac{} \cite{Bauer:2000cp,Vollinga:2005pk} to
perform the necessary algebraic manipulations. The set of integrals is
constrained by
\begin{align}
  r\in[r_{\min},r_{\max}],\qquad s\in[s_{\min},s_{\max}],
  \label{eq:rs}
\end{align}
where $r$ and $s$ are defined in \Eq{eq:rs-value}.
$r_{\max}$, $r_{\min}$, $s_{\max}$ and $s_{\min}$ are user-defined values which
control the set of seed integrals for which equations are generated. The
integrals $T(d,\bm{a})$ outside the interval limits which may be generated by
applying IBP- and LI-relations to the seeds are called auxiliary integrals.  The
rank of the system of equations $\CG$ is always smaller than the number of
different unknowns $T(d,\bm{a})$ in the system.  The goal of the Laporta
reduction is to find a representation of all the seed integrals in terms of a
small set of independent integrals, the so called master integrals. In practical
applications only the master integrals are needed to be calculated by means of
analytic or numeric algorithms since all other integrals appearing in the
calculation can be expressed as linear combinations of them. In the following we
present some implementation details of the reduction algorithm within \kira{}.

\subsection{Ordering of integrals and equations}
\label{ordering}

\paragraph{Ordering of the Integrals:}

To define an order on the integrals, in \kira{} each integral $T(d,\bm{a})$ is
represented as a list of integer numbers $\{T,S,r,s,\bm{s},\bm{r}\}$, where $T$
represents the topology, $S$ is the sector id, $r$ the sum of positive indices
in $\bm{a}$ (\Eq{eq:rs-value}), $s$ the negative sum of negative indices
(\Eq{eq:rs-value}), $\bm{s}$ the vector of non-positive indices
and $\bm{r}$ the vector of positive indices. Note that $\bm{s}$ may contain zeros. The integrals
are then ordered lexicographically with respect to $\{T,S,r,s,\bm{s},\bm{r}\}$. Integrals
which compare larger in this sense are regarded as more complicated.

\paragraph{Ordering of the Equations:}

The ordering of the integrals is used to define a (pre)order of the equations.
\kira{} represents each equation as a list of integrals $\CI$, including the
coefficients of the integrals,
\begin{align}
  0 = \sum_i c_i T(d,\bm{a}_i)
  \quad\Rightarrow\quad
  \CI = \{c_1 T_1(d,\bm{a}_1),~c_2 T_2(d,\bm{a}_{2}),~\ldots\}.
\end{align}
The integrals within each list are ordered in descending order,
i.e.\ $T_i(d,\bm{a}_i)>T_j(d,\bm{a}_j)$ for $i<j$. The first integral in each
equation is thus the most complicated one which appears in it and it serves as a
natural first criterion for the complexity of the equation. While this is in
principle sufficient to make the reduction algorithm work, it is convenient to
add further criteria to impose an order of equations with the same most
complicated integral. As the second criterion we choose the length of the
equation, followed by the remaining integrals. I.e.\ the equations are ordered
lexicographically with respect to:
\begin{align}
  \{T_1(d,\bm{a}_1),~\mathrm{length}(\CI),~T_2(d,\bm{a}_2),~T_3(d,\bm{a}_3),~\ldots\}.
\end{align}
The system of equations $\CG$ is thus represented as an ordered list of
equations
\begin{align}
  \CE = \{\CI_1,\dots \CI_n\},
  \label{listofequations}
\end{align}
where $\CI_i$ denotes the $i$-th equation and $\CI_i\le\CI_j$ if $i<j$.

Note that this defines a total preorder on the equations rather than a total
order, because it does not take into account the coefficients. Hence, equations
which contain the same set of integrals are regarded as equally complex, even if
they are not just multiples of each other.

\subsection{Reduction procedure}
\label{reduction}

\subsubsection{Selection of linearly independent equations}
\label{PYRED}

In large systems of IBP equations it has been observed that a quite large
fraction of the equations are linearly dependent, i.e.\ these equations can be
removed from the system without affecting the solution. Given that the algebraic
manipulations of the integral coefficients involve multivariate rational
functions in the kinematic invariants and the dimension $d$ it is highly
desirable to avoid any superfluous calculations involving linearly dependent
equations. This is important both to prevent expression swell at intermediate
steps and to avoid unnecessary time consuming algebraic simplifications of the
integral coefficients when different equations are combined.

An algorithm to identify linearly dependent relations based on modular
arithmetic has been presented in Ref.~\cite{Kant:2013vta} together
with an implementation in the computer program \ice{}. 
To our knowledge, the application of modular techniques, which are
well-known in mathematics \cite{Kauers:2008zz}, to solve systems of
IBP equations was discussed in Ref.~\cite{vonManteuffel:2014ixa}.
Instead of \ice{}, \kira{} uses
\pyred{} to identify redundant equations. \pyred{} is a \cplusplus{}
port of a component of an unpublished integral reduction framework
originally written in \python{}. It differs in two major ways from the
algorithm proposed in Ref.~\cite{Kant:2013vta}. First, by using larger
prime numbers in the modular arithmetic (and optionally arrays of
finite integers as coefficients), the ``Monte Carlo approach'' is
avoided. I.e.\ only a single run is required to obtain a reliable
result. Second, a variant of the Gaussian elimination algorithm is
chosen which exploits the sparsity of the system of IBP equations.

Each equation of the system is a linear combination of integrals with polynomial
or rational coefficients. In the first step, \pyred{} maps all coefficients to
a finite integer field, which is defined by a (large) prime number $p$. The
required algebraic operations are defined modulo $p$. In particular, because $p$
is prime, the multiplicative inverse $x\equiv a^{-1}$ of each finite integer
$a\in\{0,\dots,p-1\}$ is guaranteed to exist and can be calculated by solving
the equation $ax=1~(\mathrm{mod}\,p)$ for $x$ by the extended Euclidean
algorithm or by modular exponentiation $x=a^{p-2}~(\mathrm{mod}\,p)$ \cite{Gathen:2013}. Numeric
implementations of the former tend to be a bit more efficient than binary
modular exponentiation. All variables in the coefficients, i.e.\ external
(Mandelstam) invariants, masses and the dimension $d$ are substituted by
pseudo-random numbers $\in\{0,\dots,p-1\}$. Operations on finite integers are of
constant complexity, i.e.\ the time for such an operation does not depend on the
complexity of the original rational function.

The equations in the system are ordered as described in section \ref{ordering}.
The forward elimination\footnote{Forward and backward elimination
  refer to the two major steps in the algorithm to
  bring a matrix into diagonal form. In the forward elimination the
  upper triangle representation is achieved. In the following backward
  elimination the diagonal form is achieved by working in the opposite
  direction.} is then performed as follows. For each equation in the
system, substitute all previous equations in order of descending complexity. By
this procedure, the sparsity of the system is retained to a large degree,
whereas this would not be the case in the standard Gaussian elimination. The
computational complexity of Gaussian elimination on a dense system of size $n$
is of $\mathcal{O}(n^3)$. This also holds in the case of sparse systems which
become dense in intermediate steps due to an inconvenient choice of the forward
elimination algorithm. With our algorithm, the size of equations is largely
independent of the system size, i.e.\ of $\mathcal{O}(1)$, which reduces the
complexity of the entire algorithm to $\mathcal{O}(n^2)$. 
Note that we do not
perform pivoting apart from the initial ordering of the equations, thus avoiding
the additional computational cost of a pivoting operation. For optional usage we
also implemented a forward elimination algorithm with the pivoting of
Ref.~\cite{Kant:2013vta}. For the price of drastically inferior scaling behaviour and
memory consumption in \pyred{} this may in some cases lead to a better choice of
independent equations in the sense that the following reduction steps in \kira{}
are more efficient.

It is sufficient to just perform the forward elimination to identify redundant
equations. However, we chose to perform the backward elimination by default as
well. 
This operation is computationally cheap and it comes with two advantages. First,
it allows us to extract the set of master integrals already at this stage.
Second, it allows us to trace insertions of equations down to a full reduction
in such a way, that we can extract a subsystem of equations which may be
significantly smaller than the original system, but will suffice to fully reduce
all integrals of a user-specified list.

Note that it is in principle possible to reconstruct rational
functions from finite fields (see Ref. \cite{Kauers:2008zz}). This was
discussed in the context of Feynman integral reduction in
Refs.~\cite{vonManteuffel:2014ixa} and \cite{Peraro:2016wsq}.  
A private implementation for single scale
reduction problems was described in \cite{vonManteuffel:2016xki}. For
now we do not attempt to perform such a reconstruction. However,
thanks to \pyred{}'s capability to deal with arrays of coefficients,
once a reconstruction library becomes available, its integration into
\pyred{} should be straight forward.

\subsubsection{Gauss type forward elimination}
\label{forward-elimination}
After these preparatory steps---ordering of the equations and removal of
linearly dependent relations---the reduction procedure itself is started. A
Gauss type forward elimination algorithm brings the system of equations $\CG$
into triangular form. The list of equations $\CE$ may contain equations which
share the most complicated integral. First, \kira{} collects equations $\CI_i$,
which share the most complicated integral in lists of equations $\CI^{(k)}$ with
elements $\CI^{(k)}_\ell$, $\ell=1\dots m_k$, where $m_k$ is the number of
equations in the list. The original system of equations $\CE$ is thus replaced
by $\CE^*$,
\begin{align}
  \CE^* = \{\CI^{(1)},\dots, \CI^{(n)}\},
  \label{systemofsublists}
\end{align}
where the sub-lists in $\CE^*$ are ordered according to the most complicated
integral within each sub-list, and the sub-lists $\CI^{(k)}$ themselves are
ordered according to section \ref{ordering}.

To produce the upper-right triangular form the following algorithm is applied.
\begin{verse}
  repeat\\
  $\quad$ flag $=$ true\\
  $\quad$ for $\CI^{(i)}$ in $\CE^{*}$:\\
  $\quad\quad$ if $m_i>1$:\\
  $\quad\quad\quad$ flag $=$ false\\
  $\quad\quad\quad$ for $j=2,~m_i$:\\
  $\quad\quad\quad\quad$ substitute $\CI^{(i)}_1\to\CI^{(i)}_j$\\
  $\quad$ collect $\CE^*$\\
  until (flag)
\end{verse}
In this notation, ``substitute $A\to B$'' means that equation $A$ is used to
eliminate the integral $A_1$ in equation $B$. ``collect $\CE^*$'' means that the
equations in $\CE^*$ are rearranged into sorted sub-lists
of equal most complicated integrals as in \Eq{systemofsublists}. When the
algorithm terminates, $\CE^*$ is composed of lists containing only a single
equation each and an upper-right triangular form is achieved. Note that in the
implementation presented here all relations for the auxiliary integrals
(equations in which the most complicated integral is an auxiliary integral) are
dropped.

\subsubsection{Back substitution}
\label{back-substitution}
Having reached the upper-right triangular form of the system the aim
of the back substitution is to express all
seed integrals as linear combinations of master integrals. This
is done using the following algorithm.
\begin{verse}
  $\CE =$ join sub-lists $\CE^*$\\
  for $i=1$, length $\CE$:\\
  $\quad$ for $k=1,~i-1$:\\
  $\quad\quad$ substitute $\CI_k\to\CI_i$\\
  $\quad$ end for\\
  end for
\end{verse}
Here, ``join sub-lists $\CE^*$'' converts the list of lists $\CE^*$, where each
sub-list has length one, into a plain list of equations as in
\Eq{listofequations}. When the algorithm terminates all integrals are expressed
in terms of master integrals.%
\footnote{
  Strictly speaking the algorithm only guarantees to express integrals within
  the chosen seed in terms of a smaller set of integrals. Equations involving
  integrals beyond the seed (and in some cases at the edge of the seed) will
  still contain linearly dependent integrals.
}
In most applications the back substitution is the most time consuming step in
the reduction procedure.  For multi-scale problems \kira{} employs a special
strategy to perform the substitution. In a first step the back substitution is
performed and the result is sorted again with respect to the master integrals.
However, the coefficients are kept as a list and are not yet combined in one
coefficient. A naive combination of these coefficients in one step is very time
consuming since large intermediate expressions are generated. This is
avoided by combining the coefficients pairwise as follows:
\begin{enumerate}
  \item Gather all coefficients in a list, sorted by the length of the
        coefficients.
  \item A free Fermat worker process takes the two shortest coefficients from
        the list and combines them. This is done by all worker processes
        in parallel.
  \item Whenever two coefficients have been combined, add the result back to the
        list, keeping it sorted at all times. Proceed with step two unless the
        length of the list is one.
\end{enumerate}
The list in point 1.\ usually contains hundreds of coefficients, so that there
is plenty of potential for parallelization. Whenever a worker process becomes
idle, because the list contains only one item, it is assigned to another
coefficient, either from a different master integral or from a different
equation.

\subsubsection{Simplifying multivariate rational functions with Fermat}
\label{Fermat}

The integral coefficients in the Laporta system of equations are in general high
degree multivariate rational functions of the kinematic invariants, the masses
of the massive propagators, and the dimension $d$. In intermediate steps the
expressions tend to grow very large and must be simplified regularly. To
simplify the coefficients, \kira{} makes use of \fermat{} \cite{Fermat}. The
expressions are passed to \fermat{} which then performs a simplification by canceling multivariate 
rational functions. The communication between \kira{} and \fermat{} is established 
via \gatetofermat{} \cite{FLINK} which connects the two programs using UNIX pipes.

\subsubsection{Storing intermediate results using the database \sqlite{}}
As mentioned before, the integral coefficients tend to grow during the reduction
procedure. At a certain point the main memory of the computer may no longer
suffice to store the entire system of equations. Therefore \kira{} writes
equations to the hard disk and deletes them from the main memory if they are no
more needed to solve the remaining system. When writing the equations no longer
used for the back substitution to the hard disk the equations are no longer
ordered. To handle the stored equations in an efficient way and keep the
equations ordered on disk, an \sqlite{}\cite{SQLite3} database is used. \sqlite{} provides a
self-contained light-weighted SQL database. The library takes also care to order
equations encountered in subsequent write operations according to the Laporta
order described in the section \ref{ordering}.

\section{Installation}
\subsection{Prerequisites}
\label{prerequisites}
\kira{} is distributed under the terms of the:
\begin{verse}
   \texttt{ GNU General Public License, version 3 or later as published}\\ 
   \texttt{ by the Free Software Foundation.}
\end{verse}

\kira{} uses the libraries \ginac{} \cite{Bauer:2000cp, Vollinga:2005pk} (which
itself requires \cln{} \cite{CLN}), \yamlcpp{} \cite{YAML}, and \zlib{}
\cite{ZLIB}. These must be installed before compiling \kira{}. In addition
\kira{} requires the program \fermat{} \cite{Fermat}.

\subsection{Compiling and installing Kira}
\label{building Kira}

The most recent version of \kira{} is available for download from
\begin{verse}
  \texttt{https://www.physik.hu-berlin.de/de/pep/tools}
\end{verse}
as a compressed archive \texttt{kira-<version>.tar.gz}, where \texttt{<version>} is a
placeholder for the version number. Uncompress the package and change into the
extracted directory with
\begin{verse}
  \texttt{tar -xf kira-version.tar.gz}\\
  \texttt{cd kira-<version>}
\end{verse}
and configure, build and install \kira{} with the following commands.
\begin{verse}
  \texttt{%
    ./configure --prefix=/path/to/install\\
    make\\
    make install
  }
\end{verse}
The \texttt{--prefix} option specifies the installation directory. If \yamlcpp{} or
\ginac{}, or \cln{} are not found during \texttt{configure}, e.g.\ because they were not
installed via the package manager, the paths to the header files and to the
libraries must be specified through environment variables. As usual this can be
achieved by setting \texttt{CPATH} and \texttt{LD\_LIBRARY\_PATH}, respectively, or by
setting (in a Bourne compatible shell) e.g.
\begin{verse}
  \texttt{%
    export GINAC\_LIBS="-L/path/to/ginac/lib -lginac"\\
    export GINAC\_CFLAGS="-I/path/to/ginac/include"
  }
\end{verse}
if \ginac{} is installed in \texttt{/path/to/ginac}. The corresponding environment
variables for \yamlcpp{} are
\begin{verse}
  \texttt{YAML\_CPP\_CFLAGS}\quad and\quad \texttt{YAML\_CPP\_LIBS}
\end{verse}
and
\begin{verse}
  \texttt{CLN\_CFLAGS}\quad and\quad \texttt{CLN\_LIBS}
\end{verse}
for \cln{}. Since \ginac{}, \cln{} and \yamlcpp{} are linked dynamically the
paths to the
shared libraries must be set explicitly (if not installed in a standard
location) e.g.\ with
\begin{verse}
  \texttt{export LD\_LIBRARY\_PATH=/path/to/ginac/lib:\$LD\_LIBRARY\_PATH}
\end{verse}
for the \ginac{} shared library if it is located in \texttt{/path/to/ginac/lib}.
Finally, \kira{} can be started with
\begin{verse}
  \texttt{/path/to/install/bin/kira -h}
\end{verse}
or just
\begin{verse}
  \texttt{kira -h}
\end{verse}
if \texttt{/path/to/install/bin} has been added to the environment variable
\texttt{PATH}. This will print out a brief description how to use \kira{}
together with a list of the supported command line options.

\section{Kira usage}
\label{using Kira}

\kira{} uses \yaml{} files to specify the input and control the
execution in a format which is largely compatible with
\reduzetwo{}.  As far as the main tasks are concerned,
\kira{} can read and execute input files prepared for
\reduzetwo{}. Options viable in \reduzetwo{} not supported in
\kira{} are ignored\footnote{A list of the \reduzetwo{} options
recognized by \kira{} is given at the end of this chapter.}.
A corresponding message will be shown at
start-up.  The usage of \kira{} is best illustrated with an example.
\begin{figure}[htpb]
\setlength{\abovecaptionskip}{-10pt plus 0pt minus 0pt}
  \begin{center}
    \includegraphics{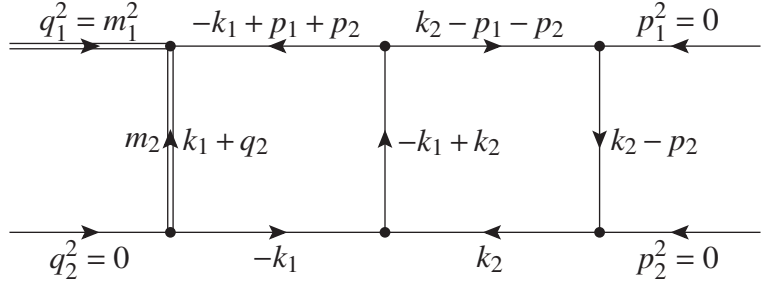}
  \end{center}
  \caption{Planar double box with one massive propagator
    and one massive external momentum (double lines). All external
    momenta  are counted ingoing.
    Momentum conservation reads $q_{1}+q_{2}+p_{1}+p_{2}=0$.}
  \label{fig:diagram topo7m}
\end{figure}
Fig.~\ref{fig:diagram topo7m} shows a double box topology as it occurs
for example in the NNLO QCD corrections to $t$-channel single
top-quark production \cite{Assadsolimani:2014oga,Brucherseifer:2014ama}.
To start a reduction \kira{} requires certain
configuration files specifying the topologies as well as kinematic
relations. In addition, a job file describing the tasks to be performed
by \kira{} is necessary.

In both cases the information is encoded in \yaml{} files.
Comments in \yaml{} files are introduced using the \texttt{\#}
sign. \yaml{} allows one to store lists and associative
lists in an easy way. In the former case the list elements are either specified
one in
a line starting with \texttt{-} in so-called \textit{block format}
or in \textit{inline format} encapsulated in square
brackets \texttt{[]}. Evidently, it is also possible to create lists
of lists. In case of associative lists a colon is used to separate a key-value pair.
An example of this notation may look like
\begin{verbatim}
momenta:
  - k1
  - k2
  - k3
loop_momenta: [l1, l2]
\end{verbatim}
Note that \yaml{} uses indentation for scoping, where only
spaces but no tabs are allowed.  The first 4 lines in the
above example define \texttt{momenta} as a list of the three momenta
\texttt{k1,k2,k3} using the block format. Similar the fifth line
declares \texttt{loop\_momenta} as list of the two momenta
\texttt{l1,l2} using the inline format.

The double box diagram shown in \Fig{fig:diagram topo7m} has
$L=2$ loop momenta and $E=3$ independent external momenta. We use
$k_{1},\;k_{2}$ to denote the two loop momenta and $q_{2},\;p_{1}$ and
$p_{2}$ for the three external momenta.
Momentum conservation is used to eliminate the fourth external
momentum $q_{1}$. In total we can
thus form $N=9$ independent scalar products involving the loop momenta
$k_1$ and $k_2$.
The scalar integral, which is associated with the
Feynman diagram shown in \Fig{fig:diagram topo7m}, is given by
\begin{align}
  T(\bm{a})=T(a_{1},\dots,a_{9})
  =\int
    d^{d}k_{1}d^{d}k_{2} \,\,\prod\limits^{9}_{j=1}\frac{1}{P_{j}^{a_{j}}},
  \label{eq:Beispiel1}
\end{align}
with
\begin{eqnarray}
  P_{1} = (-k_{1})^{2},\quad P_{2}=(k_{2})^{2},\quad
  P_{3}=(-k_{1}+k_{2})^{2},\quad
  P_{4}=(k_{1}+q_{2})^{2}-m_{2}^{2},\quad
  P_{5}=(k_{2}-p_{2})^{2},\nonumber \\
  P_{6}=(-k_{1}+p_{1}+p_{2})^{2},\quad
  P_{7}=(k_{2}-p_{1}-p_{2})^{2},\quad
  P_{8}=(k_{1}-p_{2})^{2},\quad
  P_{9}=(k_{2}-q_{2})^{2}.
  \label{eq:topo7}
\end{eqnarray}
The propagators $P_{1},\dots,P_{7}$ are associated with the $7$
internal lines, while the propagators $P_{8}$ and $P_{9}$ are
auxiliary propagators.
\kira{} uses the file \texttt{integralfamilies.yaml} located in the sub
directory
\texttt{config} of the working directory to provide the information
about the topology. Note that this file can contain more than one topology
which are distinguished by different names. For the example above, the file may look as follows.
\begin{verbatim}
#config/integralfamilies.yaml
integralfamilies:
  - name: "topo7"
    loop_momenta: [k1, k2]
    propagators:
      - ["-k1", 0]
      - ["k2", 0]
      - ["-k1+k2", 0]
      - ["k1+q2", m2^2]
      - ["k2-p2", 0]
      - ["-k1+p1+p2", 0]
      - ["k2-p1-p2", 0]
      - ["k1-p2", 0]
      - ["k2-q2", 0]
\end{verbatim}
Since \kira{} can reduce several topologies in one run, the keyword
\texttt{name} allows one to specify a name for each topology which can be
used in other files to identify the topology and control the reduction
to be done with \kira{}. The keyword \texttt{loop\_momenta} is used to
distinguish the loop momenta from the external momenta.
The keyword
\texttt{propagators} is followed by a list of the propagators. For
each propagator $P_i$ the momentum flow $l_i$ and the mass $m_i$ is
specified in the format \texttt{["l\_i", m\_i\^{}2]}.  To provide
information concerning kinematic relations like for example the
masses of the external particles or the independent invariants which
should be used to express the scalar products of external momenta, the
\yaml{} file \texttt{kinematics.yaml} is used. Like
\texttt{integralfamilies.yaml} it must also be located in the
subdirectory \texttt{config}. For the example shown in
\Fig{fig:diagram topo7m} the file may have the following form.
\begin{verbatim}
#config/kinematics.yaml
kinematics:
  incoming_momenta: [q1,q2,p1,p2]
  outgoing_momenta: []
  momentum_conservation: [q1,-p1-p2-q2]
  kinematic_invariants:
    - [s,  2]
    - [t,  2]
    - [m1, 1]
    - [m2, 1]
  scalarproduct_rules:
     - [[p1,p1],  0]
     - [[p2,p2],  0]
     - [[q2,q2],  0]
     - [[p1+p2, p1+p2], s]
     - [[p2-q2, p2-q2], t]
     - [[p1-q2, p1-q2], s-t-m1^2]
  symbol_to_replace_by_one: m1
\end{verbatim}
The keywords \texttt{incoming\_momenta} and \texttt{outgoing\_momenta} are used to
specify which external momenta are counted ingoing and which outgoing.
Since in the above example we decided  to count all momenta ingoing an
empty list is provided for the outgoing momenta. The keyword \texttt{
  momentum\_conservation} is used to specify which momentum can be
removed by applying momentum conservation. Here, the momentum $q_1$ is
replaced using
\begin{equation}
  q_1 = -p_1 - p_2 -q_2.
\end{equation}
The variables which are
used to denote the independent invariants are listed in the section
introduced by the keyword \texttt{kinematic\_invariants}. For each
variable its name and its mass dimension is provided in a list with
two elements.
The section started with the keyword \texttt{scalarproduct\_rules}
expresses the scalar products of external momenta in terms of the
invariants. To simplify the calculation it is very often useful---if
not crucial---to reduce the number of independent mass scales by one
by expressing all masses and scalar products in units of one freely
chosen invariant.
In these units the corresponding invariant is fixed to the
numerical value one. The number of variables to be treated
symbolically is thus reduced by one. To achieve this, the keyword
\texttt{symbol\_to\_replace\_by\_one} is used.  In the above example
the mass $m_1$ is set to one.

As usual we assume that dimensional
regularization is used to regulate divergent integrals.
\kira{} uses the symbol \texttt{d} to specify the space
time dimension. The symbol \texttt{d} is thus reserved and should not
be used to describe momenta or invariants.

Having provided the information about the integral topology and the
kinematics an additional \yaml{} file is used to control \kira{}. The
following lines show a minimal example:
\begin{verbatim}
#jobs1.yaml
jobs:
 - reduce_sectors:
    sector_selection:
      select_recursively:
        - [topo7,127]
    identities:
      ibp:
        - {r: [t,7], s: [0,1]}
\end{verbatim}
Since the job file is provided as command line argument to \kira{}, the name can
be freely chosen by the user. This file specifies how to prepare
and run the reduction. In the first stage at runtime, the IBP and LI equations are derived in
symbolic form and symmetry relations for the respective sectors are
prepared. To calculate the IBP and LI equations in symbolic form \kira{} uses
\ginac{}~\cite{Bauer:2000cp,Vollinga:2005pk}. Also the trivial sectors are identified.
In the second stage, the reduction is performed in four steps.
First the system of equations is generated by
evaluating the IBP and LI equations for specific powers (`seeds') of
the propagators.
In the next step, a linearly independent set of equations is chosen from the
system of equations by the \pyred{} module. In the third step, the algorithm
described in section \ref{forward-elimination} is applied
to derive the upper-right triangular form. In
the last step, the back substitution is performed as described in
section \ref{back-substitution}.
The individual steps of the reduction can be performed in separate runs.

To select the integral sectors to be reduced
the keyword \texttt{sector\_selection} is used followed by the
method to select the sectors. At the moment only
one method is available, namely
\begin{description}
\item[\texttt{select\_recursively:}] Select recursively all required
  sectors and sub-sectors to reduce the specified sector for a specific
  topology. Topology and sector are provided as a list of the form
  \texttt{[topo, sector]}.  The sector is identified using the sector
  id as defined in \Eq{eq:sectornumber}.
  It is possible to provide more than one pair (topology, sector).
\end{description}
In the above example sector 127 of topology \texttt{topo7} as
specified in \texttt{integralfamilies.yaml} together with all required
sub-sectors will be reduced.

In the setup phase \kira{} generates both the IBP as well as the LI
relations in algebraic form.
The keyword \texttt{identities} controls which seeds
should be used to generate the system of equations. Note that in
contrast to \reduze{}/\reduzetwo{} \kira{} always uses IBP and LI equations.
\begin{description}
\item[\texttt{ibp:}]
  The allowed ranges for $r$ and $s$ as defined
  in \Eq{eq:rs-value}. In the above
  example $r$ and $s$ are restricted to the range $r\in [t,7]$ and
  $s\in [0,1]$. The variable $t$ is defined in \Eq{eq:t-definition}. \kira{}
  replaces the symbol $t$ applying \Eq{eq:t-definition} to the current
  sector. If more than one associated list specifying the range for
  $r$ and $s$ is defined the set of seeds used in the run is the union.
\end{description}

Having prepared these files the working directory should contain the following files.
\begin{verbatim}
jobs1.yaml
config
config/integralfamilies.yaml
config/kinematics.yaml
\end{verbatim}
To run the reduction \kira{} is started with the file name of the job
file as command line argument,
\begin{verse}
  \texttt{kira jobs1.yaml}
\end{verse}

Note that to use \fermat{}, the path to the executable must be configured
through the environment variable \texttt{FERMATPATH}, e.g.\ with
\begin{verbatim}
export FERMATPATH="/path/to/Fermat/fermat_executable"
\end{verbatim}
After a successful run, \kira{} writes out the master integrals as
identified during the reduction. In addition, for all topologies
specified in the file \texttt{integralfamilies.yaml} the result of the
reduction is stored topology wise in subdirectories of the directory
\texttt{results}. The sector mappings and the trivial sectors of each topology
are stored topology wise in subdirectories of the directory
\texttt{sectormappings}. \texttt{sectormappings} and \texttt{results} are
located in the working directory.  In the above
example only one topology is reduced and the directories \texttt{results} and
\texttt{sectormappings} contain only one subdirectory \texttt{topo7}.
The \texttt{results} directory contains for each topology the following files:
\begin{description}
\item[\texttt{id2int}] The definition of the scalar integrals.
\item[\texttt{kira}] The result of the reduction.
\item[\texttt{kira.db}] An \sqlite{} database storing the result of
  the reduction. The data can be inspected using the program \sqlite{}.
\item[\texttt{masters}] The potential master integrals as identified
  through the numerical reduction.
\item[\texttt{masters.final}] The potential master integrals as identified
  at the end of the reduction.
\end{description}
The \texttt{sectormappings} directory contains for each topology the following
files:
\begin{description}
\item[\texttt{IBP}] The IBP equations in symbolic form.
\item[\texttt{LI}] The LI equations in symbolic form.
\item[\texttt{nonTrivialSector}] The list of non-zero sectors. The
  second number counts the number of propagators.
\item[\texttt{sectorRelations}] Relations between sectors as
  determined by \kira{}.
\item[\texttt{sectorSymmetries}] Symmetries relating different
  integrals as determined by \kira{}
\item[\texttt{topology\_ordering}] The order of the topologies as
  specified by the user.
\item[\texttt{trivialsector}] The list of zero sectors.
\end{description}

For the example discussed here, the result of the reduction
for \texttt{topo7} as stored in the file \kira{} may look:
\begin{verbatim}
- Eq: 
  - [7697655529472,0,14,0,2,"1"]
  - [7696581394432,0,14,0,2,"(2*s+d-2)/d"]
- Eq: 
  - [7697655791616,0,14,0,1,"1"]
- Eq: 
  - [7697655267328,0,14,0,2,"1"]
  - [7696581394432,0,14,0,2,"((-d+2)*t)/d"]
...
\end{verbatim}
Each equation is started with the keyword \texttt{-Eq:} and contains a
list of integrals appearing in the equation. In the example only the
first few lines of the output file are shown. The first entry of each
list denotes the left hand side of the equation---the integral which
is expressed in terms of the master integrals. The entries
in the square bracket denote 
\begin{itemize}
\item[--] the ID of the integral, 
\item[--]an integer specifying whether the integrals is a seed
  integral (0) or an auxiliary integral (-1)%
  \footnote{
    This field is mainly used for
    debugging. In the final result the entry should always be zero.
    }, 
\item[--] the variable $S$ as defined in \Eq{eq:sectornumber}, 
\item[--] the topology $T$, 
\item[--] the length of the equation (=total number of integrals in
  the equation),
\item[--] and the algebraic coefficient of the corresponding integral.
\end{itemize}
To
reduce the memory consumption during the run all integrals are mapped
to an integer used to uniquely identify the integral. The definition
of the integral ID's is stored in the file
\texttt{results/topo7/id2int}. The following lines show an example:
\begin{verbatim}
- [7696581394432,0,1,1,1,0,0,0,0,0,14,0,3,0,0,0]
- [7697655136256,-1,1,1,1,0,0,0,0,0,14,0,3,0,1,0]
- [7697655267328,0,1,1,1,-1,0,0,0,0,14,0,3,0,1,0]
- [7697655398400,0,1,1,1,0,-1,0,0,0,14,0,3,0,1,0]
- [7697655529472,0,1,1,1,0,0,-1,0,0,14,0,3,0,1,0]
...
\end{verbatim}
The lines should be interpreted as follows:\\
\texttt{[ID,$a_1$,$a_2$,$a_3$,$a_4$,$a_5$,$a_6$,$a_7$,$a_8$,$a_9$,
  $S$,$T$,$t$,$r-t$,$s$,debug]},\\
with $S$, $T$, $r$, $s$, $t$ as defined in section 2.  Obviously, the
specification of $S$, $r$, $s$, and $t$ is redundant since these
quantities can be calculated using the information for the $a_i$.

Converted back to standard mathematical notation the first equation in
the example shown above reads:
\begin{equation}
  topo7(d,0,1,1,1,0,0,-1,0,0)=
  \frac{1}{d}(2\,s+d-2)
  topo7(d,0,1,1,1,0,0,0,0,0).
\end{equation}
If an equation contains only one integral, the right hand side of the
equation and thus the integral is zero.
If a seed integral generated in the reduction does not appear in the
output file this integral is also zero.

The following file illustrates an example in which specific tasks are
to be performed.
\begin{verbatim}
#jobs2.yaml
jobs:
 - reduce_sectors:
    sector_selection:
      select_recursively:
        - [topo7,127]
    identities:
      ibp:
        - {r: [t,7],s: [0,1]}
    run_symmetries: true
    run_initiate: true
    run_pyred: true
    run_triangular: true
    run_back_substitution: true
\end{verbatim}
Note that the two examples \texttt{jobs1.yaml} and \texttt{jobs2.yaml}
perform the same tasks. The second example is given to illustrate how
specific tasks can be started manually using options starting
with \texttt{run\_}.  Starting individual tasks can also be used to
resume a reduction stopped at an intermediate state.
\begin{description}
\item[\texttt{run\_symmetries:}]
  This option will only prepare the reduction. In particular, the IBP and LI
  equations are derived in symbolic form. Symmetry relations for the respective
  sector are prepared and trivial sectors are determined.
\item[\texttt{run\_initiate:}] generate seeds in the allowed
  range and applies the IBP and the LI equations and the symmetry
  relations. The initiated system of equations is written to the files\\
  \texttt{tmp/[topo]/SYSTEM\_[topo]\_[sector\_id]}.\\
  The square brackets
  \texttt{[topo]} and \texttt{[sector\_id]} are replaced by the
  topology name and the sector id, eg.
  \texttt{tmp/topo7/SYSTEM\_topo7\_31}.
  Implies \texttt{run\_symmetries}.
\item[\texttt{run\_pyred:}]
  read the system of equations from the files\\
  \texttt{tmp/[topo]/SYSTEM\_[topo]\_[sector\_id]}\\
  and run \pyred{}.
  The result is a list of integers stored in the file\\
  \texttt{tmp/[topo]/independentEQS}.\\
  Each integer references an equation in the files\\
  \texttt{tmp/[topo]/SYSTEM\_[topo]\_[sector\_id]}.\\ The set of
  equations specifies the subset of linearly independent equations
  which will be solved in later steps.
\item[\texttt{run\_triangular:}]
  First the information stored in the file\\
  \texttt{tmp/[topo]/independentEQS}\\
  specifying the independent
  equations in files\\
  \texttt{tmp/[topo]/SYSTEM\_[topo]\_[sector\_id]}\\
  is read, then
  the system of linearly independent equations
  is built. Having set up the system the algorithm to achieve the upper-right
  triangular form is started and the result is written to the files\\
  \texttt{tmp/[topo]/VER\_[topo]\_[sector\_id]}.
\item[\texttt{run\_back\_substitution:}]
  Read the system of equations from the files\\
  \texttt{tmp/[topo]/VER\_[topo]\_[sector\_id]}\\
  and run the algorithm
  for the back substitution. The result is a list of rules to express the
  seed integrals through the master integrals. These relations are
  written to the file\\
  \texttt{results/[topo]/kira}.
\end{description}

\kira{} can reduce multiple integral families in the same run
if they are listed in the job file \texttt{jobs.yaml}
and are defined in the file \texttt{integralfamilies.yaml}.
The following files illustrate this.
\begin{verbatim}
#config/integralfamilies.yaml
integralfamilies:
  - name: "topo7"
    loop_momenta: [k1,k2]
    top_level_sectors: [127]
    propagators:
      - ["-k1", 0]         #1
      - ["k2", 0]          #2
      - ["-k1+k2", 0]      #3
      - ["k1+q2", "m2^2"]  #4
      - ["k2-p2", 0]       #5
      - ["-k1+p1+p2", 0]   #6
      - ["k2-p1-p2", 0]    #7
      - ["k1-p2", 0]       #8
      - ["k2-q2", 0]       #9
  - name: "topo7x"
    loop_momenta: [k1,k2]
    top_level_sectors: [508]
    propagators:
      - ["k1-p2", 0]       #8
      - ["k2-q2", 0]       #9
      - ["-k1", 0]         #1
      - ["k2", 0]          #2
      - ["k1+q2", "m2^2"]  #4
      - ["k2-p2", 0]       #5
      - ["-k1+k2", 0]      #3
      - ["k2-p1-p2", 0]    #7
      - ["-k1+p1+p2", 0]   #6
\end{verbatim}
Obviously, the two topologies differ only by the order of the
propagators. This example illustrates that \kira{} is able to map
sub-sectors of different topologies on each other and determines a
common set of master integrals for the different integral families
considered in one run.  Running the reduction for the two topologies
separately one will end up with different master integrals which
would need to be mapped on each other in a separate run. The following
example shows the job file to reduce both topologies (topo7 and topo7x)
in one run.

Symmetries are identified topology wise and integrals are preferably mapped to
topologies which have been defined earlier and to sectors with lower ID. One may
restrict symmetries such that integrals are only mapped to sub-sectors of
user-defined top-level sectors of each topology using the following option in
integralfamilies.yaml.
\begin{description}
\item[\texttt{top\_level\_sectors: [$S_1,S_2,\dots$]}] One may define
  multiple sectors for each topology, $S$ is defined in
  equation~\eqref{eq:sectornumber}.
\end{description}

The following example illustrates a couple of advanced features
to tune the reduction.
\begin{verbatim}
#jobs3.yaml
jobs:
 - reduce_sectors:
    sector_selection:
      select_recursively:
       - [topo7, 127]
       - [topo7x, 508]
    identities:
     ibp:
        - {r: [t,7], s: [0,2]}
        - {r: [t,8], s: [0,1]}
    select_integrals:
      select_mandatory_recursively:
       - [topo7, 127, 1, 2]
       - [topo7x, 508, 1, 2]
      select_mandatory_list:
       - [topo7, seeds7]
       - [topo7x, seeds7x]
    run_initiate: true
    run_pyred: true
    run_triangular: true
    run_back_substitution: true
    conditional: true
    alt_dir: "/path/to/alternative_dir"
\end{verbatim}

Instead of using all linearly independent equations
generated from the seeds within the specified boundaries one can let \pyred{} choose a smaller system which is sufficient to reduce a user-provided list of integrals. This is turned on with the keyword
\texttt{select\_integrals}
followed by the options:
\begin{description}
\item[\texttt{select\_mandatory\_recursively: [[topo,$S$,$r-t$,$s$]]}]
  For the topology \texttt{topo}
  (specified via the file \texttt{integralfamilies.yaml}) choose a set of equations
  which is sufficient to reduce
  the seed integrals bounded by ($S$,$r-t$,$s$). $S$,
  $r$, $t$, $s$ are defined in section 2. The unreduced
  integrals, which are regarded as the master integrals are written to the file
  \texttt{results/[topo]/masters}.
\item[\texttt{select\_mandatory\_list: [[topo,file]]}]
  Choose a set of equations which is sufficient to reduce
  the integrals specified in the file \texttt{file} for topology \texttt{topo}.
  The unreduced integrals, which are regarded as the master integrals are written to the file\\
  \texttt{results/[topo]/masters}.
\end{description}
Note, that \kira{} does not guarantee that all integrals within a rectangular seed selection can be reduced. When the option \texttt{select\_mandatory\_recursively} is invoked, \kira{} will print the master integrals after the numerical reduction. If an integral at the edge of the seed range appears as a master integral, this integral is unreduced. In this case the user needs to enlarge the rectangular seed selection and restart \kira{}.

The option \texttt{alt\_dir} can be used to specify a directory for the
intermediate and final results.
\begin{description}
\item[\texttt{alt\_dir: "/path/to/alternative\_dir"}]
  All temporary and result files will be saved and loaded from the directories\\
  \texttt{"/path/to/alt\_dir/tmp"}, \texttt{"/path/to/alt\_dir/results"} and \\
  \texttt{"/path/to/alt\_dir/sectormappings"}.
  If \texttt{alt\_dir} is not specified, the working directory is used.
\end{description}
\begin{description}
\item[\texttt{conditional: true}] In \kira{} the results of a backward substitution will be commited to the database \texttt{kira.db} every 10 minutes. Since \kira{} version 1.1 the backward substitution can be killed at any time. To resume the backward substitution and to load the results from a previous run the option \texttt{conditional} must be set to \texttt{true}. An option to interrupt the backward substitution gracefully will be provided in a future \kira{} version.
\end{description}

As mentioned before, each equation in the result file \kira{}
represents a rule to replace a seed integral through the master
integrals. To extract the results in a specific format usable in
computer algebra programs like \form{} \cite{Vermaseren:2000nd} or \mathematica{}
one may use the following options in the job file.

\begin{description}
\item[\texttt{- kira2form:}]
  With \texttt{target: [[topo,seeds]]} the integrals of topology \texttt{topo} listed in the file \texttt{seeds}
  will be translated into a \form{} readable file:\\
  \texttt{results/topo/kira\_seeds.inc}.\\
  With \texttt{target: [[topo,$S$,$r-t$,$s$]]} it is possible to reconstruct all integrals for the topology \texttt{topo}
  (specified via the file \texttt{integralfamilies.yaml}), which are bounded by ($S$,$r-t$,$s$). $S$, $r$, $t$, $s$ are defined in section 2. The result will be written into a \form{} readable file:  \texttt{results/topo/kira\_$S$\_$(r-t)$\_$s$.inc}.\\
  If during the reduction the option \texttt{alt\_dir: "/path/..."}
  was used, then the option \texttt{alt\_dir: "/path/..."} is mandatory. \kira{} will look for the results of the reduction
  in the directory specified via the option \texttt{alt\_dir}.
  \item[\texttt{- kira2math:}]
  This option is similar to the option \texttt{- kira2form:}. Here the results will be written into \mathematica{} readable files ending with \texttt{.m}
\end{description}
By default, the dependence of coefficients in the symbol which was replaced by
one during the reduction with the option \texttt{symbol\_to\_replace\_by\_one}
will not be reconstructed. This can be activated with the option
\texttt{reconstruct\_mass: true}. An example of a file containing the seed
integrals is shown here.
\begin{verbatim}
#seeds7
- [0,1,1,1,-1,0,0,0,0]
- [0,1,1,1,0,0,-1,0,0]
- [1,1,1,1,1,0,0,0,0]
- [1,1,1,1,2,0,0,0,0]
\end{verbatim}
A job file extracting various identities in \form{} and \mathematica{}
readable form may look like
\begin{verbatim}
#export.yaml
jobs:
 - kira2math:
   target:
     - [topo7x,seeds7x]
     - [topo7,seeds7]
     - [topo7,127,1,2]
   reconstruct_mass: true
 - kira2form:
   target:
     - [topo7x,seeds7x]
     - [topo7,seeds7]
\end{verbatim}

In case the option \texttt{kira2form} is used the file
\texttt{kira\_seeds7.inc} contains identities of the form
\begin{verbatim}
id topo7(0,1,1,1,-1,0,0,0,0) =
 + topo7(0,1,1,1,0,0,0,0,0)*(((-d+2)*t)*den(d))
;

id topo7(0,1,1,1,0,0,-1,0,0) =
 + topo7(0,1,1,1,0,0,0,0,0)*((2*s+d-2)*den(d))
;
...
\end{verbatim}
In case \texttt{kira2math} is used the file
\texttt{kira\_seeds7.m} looks like
\begin{verbatim}
{
topo7[0,1,1,1,-1,0,0,0,0] ->
 + topo7[0,1,1,1,0,0,0,0,0]*(((-d+2)*t)/d)
,
topo7[0,1,1,1,0,0,-1,0,0] ->
 + topo7[0,1,1,1,0,0,0,0,0]*(((d-2)*m1^2+2*s)/d)
,
...}
\end{verbatim}
The \mathematica{} readable format can be included in \mathematica{} using
the command
\begin{verbatim}
  rule = Get["results/topo7/kira_seeds7.m"];
\end{verbatim}
For the above example a \mathematica{} session may look like this
\begin{verbatim}
In[1]:= rule = Get["results/topo7/kira_seeds7.m"];

In[2]:= topo7[0,1,1,1,0,0,-1,0,0] /. rule

                    2
        ((-2 + d) m1  + 2 s) topo7[0, 1, 1, 1, 0, 0, 0, 0, 0]
Out[2]= -----------------------------------------------------
                                  d
\end{verbatim}
In this example the mass $m_{1}$ was reconstructed.

In addition to the options which can be specified in the configuration
files the following command line arguments are recognized by \kira{}.
\begin{description}
\item[\texttt{--version}] Print out the current \kira{} version.
\item[\texttt{--help}] Print out a brief description of the command
  line arguments and how to use \kira{}.
\item[\texttt{--silent}] Suppresses the output to the
  screen during the run. Note that the log file \texttt{kira.log} is still written.
\item[\texttt{--parallel=}\textit{n}] Run
  $n$ instances of \fermat{} in parallel. During the back
  substitution significant runtime is spent for the algebraic
  simplification of the integral coefficients. Performing this step in
  parallel can lead to a significant speed-up. In the current \kira{}
  version the maximal number of parallel tasks is unlimited. The
  value set by the user should not exceed the number of processor
  cores available. Also the generation of the
  IBP and LI equations and the algorithm to build the
  upper-right triangular form are run in parallel.
\item[\texttt{--algebra}] For multi-scale problems the integral
  coefficients tend to become rather large. In this case the option
  \texttt{--algebra} might be useful. This enables a
  modified algorithm for the back substitution and in particular the
  sorting algorithm described in section \ref{back-substitution}.
\end{description}

The following \reduzetwo{} options are supported in \kira{}:
\begin{verbatim}
jobs:
  - reduce_sectors:
    sector_selection:
      select_recursively:
    identities:
      ibp:
integralfamilies:
  - name:
    loop_momenta:
    propagators:
    cut_propagators:
kinematics:
  incoming_momenta:
  outgoing_momenta:
  momentum_conservation:
  kinematic_invariants:
  scalarproduct_rules:
  symbol_to_replace_by_one:
\end{verbatim}

\section{Benchmarks}
\label{benchmark}
\begin{figure}[htpb]
\setlength{\abovecaptionskip}{-10pt plus 0pt minus 0pt}
  \begin{center}
    \includegraphics{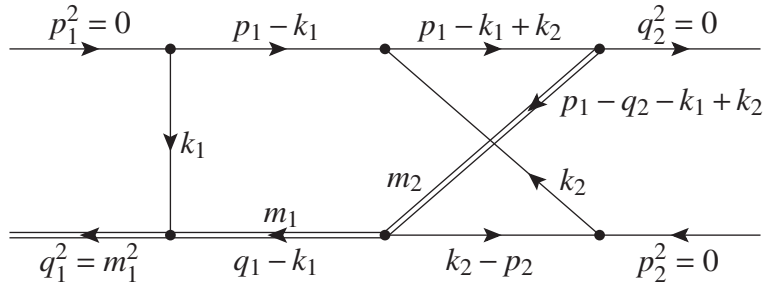}
  \end{center}
  \caption{\texttt{topo4} is a non planar double box with two massive
    propagators and one massive external momentum. Momentum
    conservation reads $q_{1}=p_{1}+p_{2}-q_{2}$.}
  \label{fig:diagram topo4}
\end{figure}

\begin{figure}[htpb]
\setlength{\abovecaptionskip}{-10pt plus 0pt minus 0pt}
  \begin{center}
    \includegraphics{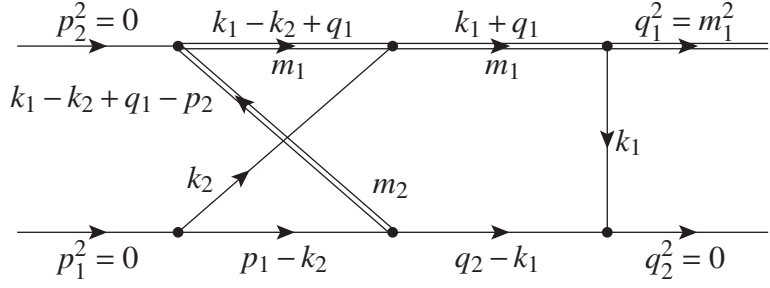}
  \end{center}
  \caption{\texttt{topo5} is a non planar double box with three massive
    propagators and one massive external momentum. The momentum
    conservation reads $q_{1}=p_{1}+p_{2}-q_{2}$.}
  \label{fig:diagram topo5}
\end{figure}

To benchmark the performance of our implementation we study three
examples, occurring in the evaluation of NNLO corrections to
$t$-channel single top-quark production.  The first example is the
planar double box \texttt{topo7} shown in \Fig{fig:diagram topo7m}.
The second example is a non planar topology \texttt{topo4} shown in
\Fig{fig:diagram topo4}. The integral associated with
\texttt{topo4} is given by \Eq{eq:Beispiel1}, with the following
definition of the propagators:
\begin{eqnarray*}
  &P_{1}=k_1^2, \quad
  P_{2}=k_2^2, \quad
  P_{3}=(p_{1}-k_{1})^2, \quad
  P_{4}=(p_{2}-k_{2})^2, \quad
  P_{5}=(p_{1}+p_{2}-q_{2}-k_{1})^2-m_1^2, \\
  &P_{6}=(p_{1}-k_{1}+k_{2})^2, \quad
  P_{7}=(p_{1}-q_{2}-k_{1}+k_{2})^2-m_2^2, \quad
  P_{8}=(k_{1}-q_{2})^2, \\
  &P_{9}=(k_{2}-p_{1}-p_{2})^2.
\end{eqnarray*}

The third example is the non planar topology \texttt{topo5} shown in
\Fig{fig:diagram topo5}. This turns out to be the most
complicated topology in single top-quark production at NNLO.
The integral associated with \texttt{topo5} is again given by
\Eq{eq:Beispiel1}, with the following propagators,
\begin{eqnarray*}
  &P_{1}=k_1^2, \quad
  P_{2}=k_2^2, \quad
  P_{3}=(q_{2}-k_{1})^2, \quad
  P_{4}=(p_{1}-k_{2})^2, \quad
  P_{5}=(q_{1}+k_{1})^2-m_{1}^2, \\
  &P_{6}=(q_{1}+k_{1}-k_{2})^2-m_1^2, \quad
  P_{7}=(-p_{2}+q_{1}+k_{1}-k_{2})^2-m_2^2, \quad
  P_{8}=(k_{1}-p_{1})^2, \\
  &P_{9}=(k_{2}-q_{2}-p_{2})^2.
\end{eqnarray*}
In both cases the propagators $P_{1},\dots,P_{7}$ are associated with
the $7$ internal lines, while the propagators $P_{8}$ and $P_{9}$ are
auxiliary propagators.

\begin{table}[htpb]
\caption{The runtime used by \kira{} to reduce topology \texttt{topo7} as
  defined in \Eq{eq:Beispiel1} and \Eq{eq:topo7}. The parameter $s$ describes
  the total power of propagators occurring in the numerator. $r_{\max}$ is set
  to 7. In addition, we also give the time $\bm{T_\pyred}$ used by the \pyred{}
  module within \kira{} to identify the linearly dependent equations. For
  comparison the runtime for the same reduction using \reduzetwo{} and \firefive{} is shown.
  \label{tab:topo7m}}
\begin{center}\renewcommand{\arraystretch}{1.8}
  \begin{tabular}[\linewidth]{ccccccccc}
    \hline
    \texttt{Type}&
    $\bm{s_{\max}}$& $\bm{T_\pyred}$ & $\bm{T_\kira}$&
    $\bm{T_\reduzetwo}$&$\bm{T_\firefive}$& $\frac{\bm{T_\pyred}}{\bm{T_\kira}}$ &
    $\frac{\bm{T_\reduzetwo}}{\bm{T_\kira}}$ &
    $\frac{\bm{T_\firefive}}{\bm{T_\kira}}$\\
    \hline
    default   &1 & 1.1 s & 142 s    & 2 h    & 17 min  & 0.008 & 51  & 7.1    \\\hline
    \texttt{A}&1 & 2.6 s & 49.5 s   & -      & 6.7 min & 0.053 & -   & 8.1 \\\hline
    \texttt{B}&1 & 2.6 s & 40 s     & -      & 1 s     & 0.065 & -   & 0.025  \\\hline\hline
%
    default   &2 & 4.5 s & 664 s    & 10 h   & 87 min  & 0.007 & 54  & 7.9   \\\hline
    \texttt{A}&2 & 4.5 s & 224 s    & -      & 20.3 min& 0.02  & -   & 5.4    \\\hline
    \texttt{B}&2 & 4.5 s & 203 s    & -      & 17.6 min& 0.022 & -   & 5.2   \\\hline\hline
%
    default   &3 & 11 s  & 48.5 min & 28.4 h & 4.7 h   & 0.0004& 35  & 5.8   \\\hline
    \texttt{A}&3 & 11 s  & 14.2 min & -      & 1.47 h  & 0.013 & -   & 6.2   \\\hline
    \texttt{B}&3 & 11 s  & 10.7 min & -      & 1.3 h   & 0.017 & -   & 7.3   \\\hline\hline
%
    default   &4 & 23 s  & 4.1 h    & 4.4 d  & 13.6 h  & 0.0015& 25  & 3.3    \\\hline
    \texttt{A}&4 & 23 s  & 1.2 h    & -      & 4.5 h   & 0.005 & -   & 3.75   \\\hline
    \texttt{B}&4 & 23 s  & 35.4 min & -      & 3.9 h   & 0.01  & -   & 6.6    \\\hline
  \end{tabular}
  \end{center}
\end{table}
All benchmarks were run on compute nodes equipped with
two Intel(R) Xeon(R) E5-2680 CPUs (8 cores/CPU) clocked at
2.70\,GHz and 396 GBytes of RAM.

As described in the previous section, \kira{} offers the ability to reduce
only selected integrals using the option \texttt{select\_integrals}.
We study three different types of jobs to reduce integrals of the complexity
$s_{\max}=1,2,3,4$. The jobs of type \texttt{default} do not employ
the option \texttt{select\_integrals}. Type \texttt{A} jobs use the option
\texttt{select\_mandatory\_recursively: [[topo,127,0,$s_{\max}$]]}, where
\texttt{topo} is replaced by \texttt{topo4} or \texttt{topo7}. This option
forces \kira{} to reduce integrals with $r=t$ (zero dots) and
$s=\{0,\dots,s_{\max}\}$. Finally, the jobs of type \texttt{B} use the option
\texttt{select\_mandatory\_list: [[topo,list]]}, where \texttt{topo} is again
replaced by \texttt{topo4} and \texttt{topo7} and \texttt{list} is replaced by
\texttt{list1}, \texttt{list2}, \texttt{list3} and \texttt{list4} for
$s_{\max}=1,2,3,4$, respectively:
\begin{verbatim}
#list1
- [1,1,1,1,1,1,1, 0,-1]
- [1,1,1,1,1,1,1,-1, 0]

#list2
- [1,1,1,1,1,1,1, 0,-2]
- [1,1,1,1,1,1,1,-2, 0]
- [1,1,1,1,1,1,1,-1,-1]

#list3
- [1,1,1,1,1,1,1, 0,-3]
- [1,1,1,1,1,1,1,-3, 0]
- [1,1,1,1,1,1,1,-2,-1]
- [1,1,1,1,1,1,1,-1,-2]

#list4
- [1,1,1,1,1,1,1, 0,-4]
- [1,1,1,1,1,1,1,-4, 0]
- [1,1,1,1,1,1,1,-1,-3]
- [1,1,1,1,1,1,1,-3,-1]
- [1,1,1,1,1,1,1,-2,-2]
\end{verbatim}

We start our discussion with the reduction of topology \texttt{topo7}.
In the benchmark we use the parameter $s$, counting the total power of
propagators in the numerator, to control the complexity of the
reduction. For all reductions we have checked
that \kira{}, \reduzetwo{} and \firefive{}%
\footnote{
  The used \fermat{} is 64 bit Linux version 5.25.
  For the benchmarks with \reduzetwo{} we used version 2.1.2 (MPI build) and with \firefive{} we used version 5.2.
} 
produce the same set of master integrals and that
the results for the reduced integrals agree. \Tab{tab:topo7m}
shows the runtime used by \kira{} and \reduzetwo{}. In addition, we report also
the runtime used by the \pyred{} module. While the time used in \pyred{} is
small compared to the total runtime, removing the linearly dependent
equations significantly reduces the total time required by \kira{}.
We observe that in the considered examples \kira{} is between 1--2 orders of
magnitude faster than \reduzetwo{}. For the reduction of type \texttt{default} \kira{} is up to 1 order of magnitude faster than \firefive{}, but for the reduction type \texttt{A} and type \texttt{B} \kira{} is between 1--2 orders of magnitude faster than \firefive{}.

\begin{table}[htpb]
  \caption{Same as \Tab{tab:topo7m} but for topology \texttt{topo4}.
    ($r_{\max}$ is set to 7.)
    In all reductions one mass scale is removed using the ratio
    $m_{2}^{2}=\frac{3}{14}m_{1}^{2}$.
    \reduzetwo{}, \firefive{} and \kira{} were initialized with 11 cores. \label{tab:topo4}
  }
  \begin{center}\renewcommand{\arraystretch}{1.8}
  \begin{tabular}[\linewidth]{ccccccccc}
    \hline
    \texttt{Type}&
    $\bm{s_{\max}}$& $\bm{T_\pyred}$ & $\bm{T_\kira}$&
    $\bm{T_\reduzetwo}$&$\bm{T_\firefive}$&
    $\frac{\bm{T_\pyred}}{\bm{T_\kira}}$  &
    $\frac{\bm{T_\reduzetwo}}{\bm{T_\kira}}$&
    $\frac{\bm{T_\firefive}}{\bm{T_\kira}}$\\
    \hline
    default   &1 & 2.8 s  & 90 s     & 2.1 h  & 23 min  & 0.03  & 86 & 15.3  \\\hline
    \texttt{A}&1 & 2.8 s  & 23.6 s   & -      & 19.3 min& 0.11  & -  & 49    \\\hline
    \texttt{B}&1 & 2.8 s  & 16.1 s   & -      & 1.6 s   & 0.17  & -  & 0.1   \\\hline\hline
    default   &2 & 9.8 s  & 6.6 min  & 7.2 h  & 2.3 h   & 0.02  & 65 & 21    \\\hline
    \texttt{A}&2 & 11.3 s & 167 s    &  -     & 2.2 h   & 0.07  & -  & 47    \\\hline
    \texttt{B}&2 & 11.2 s & 160 s    &  -     & 2.2 h   & 0.07  & -  & 50    \\\hline\hline
    default   &3 & 28 s   & 43 min   & 22.8 h & 7.6 h   & 0.01  & 32 & 10.6  \\\hline
    \texttt{A}&3 & 30.4 s & 539 s    & -      & 7.4 h   & 0.06  & -  & 49.4  \\\hline
    \texttt{B}&3 & 30.1 s & 444 s    & -      & 7.5 h   & 0.07  & -  & 61    \\\hline\hline
    default   &4 & 67 s   & 2.4 h    & 2.7 d  & 23.5 h  & 0.007 & 26 & 9.8   \\\hline
    \texttt{A}&4 & 70.2 s & 35.3 min & -      & 22.4 h  & 0.03  & -  & 38    \\\hline
    \texttt{B}&4 & 69.5 s & 24 min   & -      & 22.4 h  & 0.05  & -  & 56    \\\hline
  \end{tabular}
  \end{center}
\end{table}

In case of topology \texttt{topo4} the additional mass scale $m_2$
leads to a significant increase in complexity. In single top-quark
production $m_1$ corresponds to the top-quark mass and $m_2$ is
the $W$ boson mass. In Ref.~\cite{Assadsolimani:2014oga} a fixed
ratio between the two masses was used to reduce the number of
independent scales and thus the complexity of the reduction. In the
benchmark presented here we follow the same strategy and set
\begin{equation}
  m_{2}^2=\frac{3}{14}m_{1}^{2}.
\end{equation}
The runtime required for the reduction of the topology \texttt{topo4} is
given in \Tab{tab:topo4}. Since in the \pyred{} module the invariants are
replaced by integer values the runtime for this part of the reduction
is similar to the runtime observed for the topology \texttt{topo7}.
Again only a small fraction of the total runtime is required to
identify the linearly dependent equations. Even for the most
complicated reduction the required runtime is roughly a minute.
\reduzetwo{}, \firefive{} as well as \kira{} were all started with 11 cores allowing to
perform a significant part of the reduction in parallel.
The examples presented in \Tab{tab:topo7m} and \Tab{tab:topo4} require
very little memory. An amount of 4 GBytes is sufficient to run the examples.
In Tabs.~\ref{tab:topo7m} and \ref{tab:topo4} we observe that the jobs of type \texttt{A} and type \texttt{B} improve the runtime of \texttt{Kira} by a factor of 3--5 compared to the job of type \texttt{default} without selecting the integrals with the option \texttt{select\_integrals}.

\begin{table}[htpb]
  \caption{The run time $\bm{T_\pyred}$ for \pyred{} which is called
    by \kira{} is shown and compared to the total time $\bm{T_\kira}$,
    which \kira{} needed for a complete reduction of the topology topo4 and
    topo5 keeping the full mass dependence. \kira{} was initialized
    with the options
    \texttt{--algebra} and \texttt{--parallel=13}.}
  \label{tab:topo4andtopo5-with-full-mass-dep}\renewcommand{\arraystretch}{1.3}
  \begin{center}
  \begin{tabular}[\linewidth]{cccccc}
    \hline
    $\bm{\mr{Topology}}$&$\bm{r_{\max}}$&$\bm{s_{\max}}$&
    $\bm{T_\pyred}$ &
    $\bm{T_\kira}$ & $\frac{\bm{T_\pyred}}{\bm{T_\kira}}$ \\
    \hline
    topo4 & 8 & 3 & 41 s   & 14 h & 0.0008  \\
    & 7 & 4 & 130 s  & 10 h & 0.003   \\\hline
    topo5 & 8 & 2 & 94 s   & 3 d  & 0.0003  \\
    & 8 & 3 & 125 s  & 8 d  & 0.0002  \\
    & 7 & 4 & 237 s  & 7 d  & 0.0004  \\\hline
  \end{tabular}
  \end{center}
\end{table}

As a final benchmark we study the reduction of \texttt{topo4} and
\texttt{topo5} keeping the full mass dependence. The runtime required
is shown in \Tab{tab:topo4andtopo5-with-full-mass-dep}. The reduction
was done using 13 processor cores. In addition, the command line
option \texttt{--algebra} was used to reduce the time required for the
back substitution.  Comparing the results for topology \texttt{topo4} shown in
\Tab{tab:topo4} and \Tab{tab:topo4andtopo5-with-full-mass-dep}, we
observe that the additional mass scale leads to significant increase
in the total runtime. As mentioned before, the time required by
the \pyred{} module to eliminate the linearly dependent equations is only
mildly affected since this part is based on integer arithmetic.  For
\texttt{topo4} the total runtime is of the order of 10 hours while for
\texttt{topo5} the most challenging reductions take roughly one week.
Most of the time is spent on the algebraic simplifications of the
integral coefficients using \fermat{}. This is also reflected in significantly
increased memory consumption. To reproduce the results shown in
\Tab{tab:topo4andtopo5-with-full-mass-dep} about 90 GBytes of RAM is
required in \kira{} plus around 10 GBytes for each \fermat{} instance.
Again, for all reductions we have checked that \kira{} and \reduzetwo{} produce the
same set of master integrals. To compare the reduction against \reduzetwo{} we ran
\reduzetwo{} with numerical input values for the kinematics instead of symbolic
input.

The gain in performance using the option \texttt{--algebra}
is illustrated in \Tab{tab:--algebra-impact}. As expected the
improvement depends on the complexity. For the simplest case
($s_{\max}=1$) the total runtime is roughly reduced by a factor 2.
Increasing the complexity ($s_{\max}=3$) a total speed-up by roughly a
factor 4.5 is achieved.

\begin{table}[htpb]
  \caption{
    Runtime used for the reduction of
    topology topo4 keeping the full mass dependence for
    different
    complexities $s_{\max}$. $r_{\max}$ is set to 7.
    \kira{} is started with different command line options.
    }
  \label{tab:--algebra-impact}\renewcommand{\arraystretch}{1.3}
  \begin{center}
  \begin{tabular}[\linewidth]{cccc}
    \hline
    $\bm{s_{\max}}$& $\bm{T_{\mr{Back\;substitution}}}$ &
    $\bm{T_{\mr{Total}}}$ & \textbf{options} \\
    \hline
      1 & 495 s   & 543 s   & \texttt{--algebra --parallel=16}  \\
      1 & 1108 s  & 1156 s  & \texttt{--parallel=16}   \\\hline
      2 & 2920 s  & 3354 s  & \texttt{--algebra --parallel=16}  \\
      2 & 6683 s  & 7096 s  & \texttt{--parallel=16}   \\\hline
      3 & 13203 s & 13664 s & \texttt{--algebra --parallel=16}  \\
      3 & 59905 s & 60370 s & \texttt{--parallel=16}  \\\hline
  \end{tabular}
  \end{center}
\end{table}

\section{Conclusion}
\label{summary}
In this article we presented a new implementation of the Laporta algorithm to
reduce multi-loop Feynman integrals to a small set of master
integrals. Compared to previous implementations an algorithm based on
modular arithmetic is used to eliminate linearly dependent
equations from the set of IBP and LI relations. Using only the
linearly independent equations the system is brought into upper triangle
form using a straight forward Gauss elimination. For the backward substitution
an optimized procedure
delaying the expression swell of intermediate expressions has been
implemented.  Removing linearly dependent equations in combination
with the optimized back substitution leads to a significant increase
in performance when complicated topologies are reduced. Particularly
multi-scale problems benefit from these improvements. To illustrate
the mentioned features we have successfully reproduced various
reductions occurring in the calculation of the NNLO corrections to
single top-quark production. We also stress that the algorithm is not
limited to two-loop corrections but can be applied also to higher loop
reductions.

\paragraph{Acknowledgments:}
J.U. would like to thank Bas Tausk for his very
useful discussions during the early stage of this project. We wish 
to express our special thanks to Andreas von Manteuffel, Tord Riemann and 
Bas Tausk for a careful reading of the manuscript and useful comments.
The work of J.U. is supported by the research training group GRK-1504 ``Masse,
Spektrum, Symmetrie'' funded by the German research foundation (DFG) and received funding from the European Research Council (ERC) under the European Union's Horizon 2020 research and innovation programme under grant agreement no.\ 647356 (CutLoops). P.M. 
acknowledges support by the state of Baden-Württemberg through bwHPC and the 
German Research Foundation (DFG) through grant no INST 39/963-1 FUGG.

\providecommand{\href}[2]{#2}\begingroup\raggedright\endgroup

\end{document}